\documentclass[twocolumn,aps,prl,preprintnumbers,bibnotes10pt,superscriptaddress]{revtex4}

\UseRawInputEncoding

\usepackage{amsmath}
\usepackage{graphicx}
\usepackage{ulem}
\usepackage{dcolumn}
\usepackage{epsfig}
\usepackage{bm}
\usepackage{array}
\usepackage[frenchb]{babel}
\usepackage{hyperref}
\hypersetup{
colorlinks=true,
citecolor=black,
linkcolor=red,
urlcolor=black
 , bookmarks=true,
 pdfmenubar=true
}

\usepackage{hyperref}
\usepackage{graphicx}
\usepackage{amsmath}
\usepackage{amsfonts}
\usepackage{amssymb}
\usepackage{xcolor}
\usepackage{bbm}
\usepackage{calrsfs}
\usepackage{dutchcal}
\usepackage{amsthm}

\newcommand{\ket}[1]{\ensuremath{|#1\rangle}}

\newcommand{\vect}[1]{\bm{#1}}

\newcommand{\be}{\begin{equation}}
\newcommand{\ee}{\end{equation}}

\newcommand{\beq}{\begin{eqnarray}}
\newcommand{\eeq}{\end{eqnarray}}

\begin{document}

\title{A machine learning approach to Bayesian parameter estimation}
\author{Samuel Nolan}
\affiliation{QSTAR, INO-CNR and LENS, Largo Enrico Fermi 2, 50125 Firenze, Italy}

\author{Augusto Smerzi}
\affiliation{QSTAR, INO-CNR and LENS, Largo Enrico Fermi 2, 50125 Firenze, Italy}

\author{Luca Pezz$\grave{\text{e}}$}
\affiliation{QSTAR, INO-CNR and LENS, Largo Enrico Fermi 2, 50125 Firenze, Italy}

\begin{abstract}

Bayesian estimation is a powerful theoretical paradigm for the operation of
approach to parameter estimation. However, the Bayesian method for statistical inference generally suffers from demanding calibration requirements, that have so far restricted its use to systems that can be explicitly modeled.
In this theoretical study, we formulate parameter estimation as a classification task and use artificial neural networks to efficiently perform Bayesian estimation. 
We show that the network's posterior distribution is centered at the true (unknown) value of the parameter within an uncertainty given by the inverse Fisher information, representing the ultimate sensitivity limit for the given apparatus.
When only a limited number of calibration measurements are available, our machine-learning based procedure outperforms standard calibration methods. 
Our machine-learning based procedure is model independent, and is thus well suited to ``black-box sensors'', which 
lack simple explicit fitting models.
Thus, our work paves the way for Bayesian quantum sensors that can take advantage of complex non-classical quantum states and/or adaptive protocols. 
These capabilities can significantly enhance the sensitivity of future devices.
\end{abstract}

\maketitle

%\section{Introduction}

%Clarify how multuple measurements work, either
%in the intro, abstract or conclusion. Maybe in Fig 2.

Precise parameter estimation in quantum systems can revolutionize 
current technology and prompt scientific discoveries~\cite{DegenRMP2017, PezzeRMP2018}.
Prominent examples include gravitational wave detection~\cite{SchnabelPR2017,TsePRL2019, ArcenesePRL2019}, time and frequency standards in atomic clocks~\cite{LudlowRMP2015}, field sensing in magnetometers~\cite{RondinRPP2014}, inertial sensors~\cite{CroninRMP2009, BarrettPS2016} and biological imaging~\cite{TaylorPR2016}. 
As such, improving the sensitivity of quantum sensors is currently an active area of research with most work focused on the control and reduction of noise and decoherence, and on the use of non-classical probe states~\cite{PezzeRMP2018}.
Furthermore, the development of data analysis techniques to extract information encoded in complex quantum states~\cite{LanePRA1993, PezzePRL2007, OlivaresJPB2009, KrischekPRL2011, XiangNATPHOT2011, PezzeBOOK2014, LiENTROPY2018, RubioJPC2018, CiminiPRA2020} is another crucial, yet often overlooked step toward ultra-precise quantum sensing.

Among different strategies~\cite{KayBOOK, LehmannBOOK, VanTreesBOOK}, Bayesian parameter estimation (BPE) is known to be particularly efficient and versatile.
The output of BPE is a conditional 
probability distribution $P(\theta \vert \vect{\mu})$ which is interpreted as a degree of belief that the parameter $\theta$ equals the true (unknown) value $\theta_{\rm true}$, given the sequence of $m$ measurement results $\vect{\mu} = \mu_1, ..., \mu_m$ and any prior information about $\theta_{\rm true}$~\cite{PezzeBOOK2014, LiENTROPY2018}. 
BPE is free of any assumption about the probability distribution of the measurement data $\vect{\mu}$, and it can meaningfully assign a confidence interval to any result, even a single detection event ($m=1$).
As $m$ becomes large, $P(\theta \vert \boldsymbol{\mu})$ converges to a Gaussian centered at $\theta_{\rm true}$ and with a width given by the inverse Fisher information, a result which crucially holds for any probability model and all values of the parameter $\theta_{\rm true}$  \cite{LehmannBOOK, VanTreesBOOK, PezzeBOOK2014}.
Finally, BPE forms the basis of several adaptive protocols in parameter estimation~\cite{BerryPRL2000, HiggingNATURE2007,
BerniNATPHOT2015, BonatoNATNANO2015, MartinezNJP2016, WiebePRL2016, PaesaniPRL2017, SantagatiPRX2019}. 
However, performing BPE necessitates a detailed characterization of the measurement apparatus, which typically requires either modelling the sensor explicitly, or else collecting a prohibitively large amount of calibration data. 
 Although BPE has been demonstrated in single-qubit systems such as NV center magnetometrs \cite{HincksNJP2012, BonatoNATNANO2015, SantagatiPRX2019, AhronSR2019, SchwartzSR2019}, its demanding calibration requirements remain a major limitation when moving to more complex systems. For example, complex non-classical states are now routinely generated in ensembles of ultra-cold atoms \cite{PezzeRMP2018}. 
BPE using entangled states has so far only limited to some proof-of-principle investigations in few-particle systems~\cite{PezzePRL2007, KrischekPRL2011, XiangNATPHOT2011}.
To employ BPE in systems that cannot be easily modeled, methods must be developed to efficiently calibrate the device given limited data.

%%%%%%%%%%%%%%%%%%%%%%%%%%%%%%%%%%%%%%%%%%%%%%%%%%%%%%%%%%%%%
% figure 1
%%%%%%%%%%%%%%%%%%%%%%%%%%%%%%%%%%%%%%%%%%%%%%%%%%%%%%%%%%%%%

\begin{figure}[t!]
\centering
\includegraphics[width=\columnwidth]{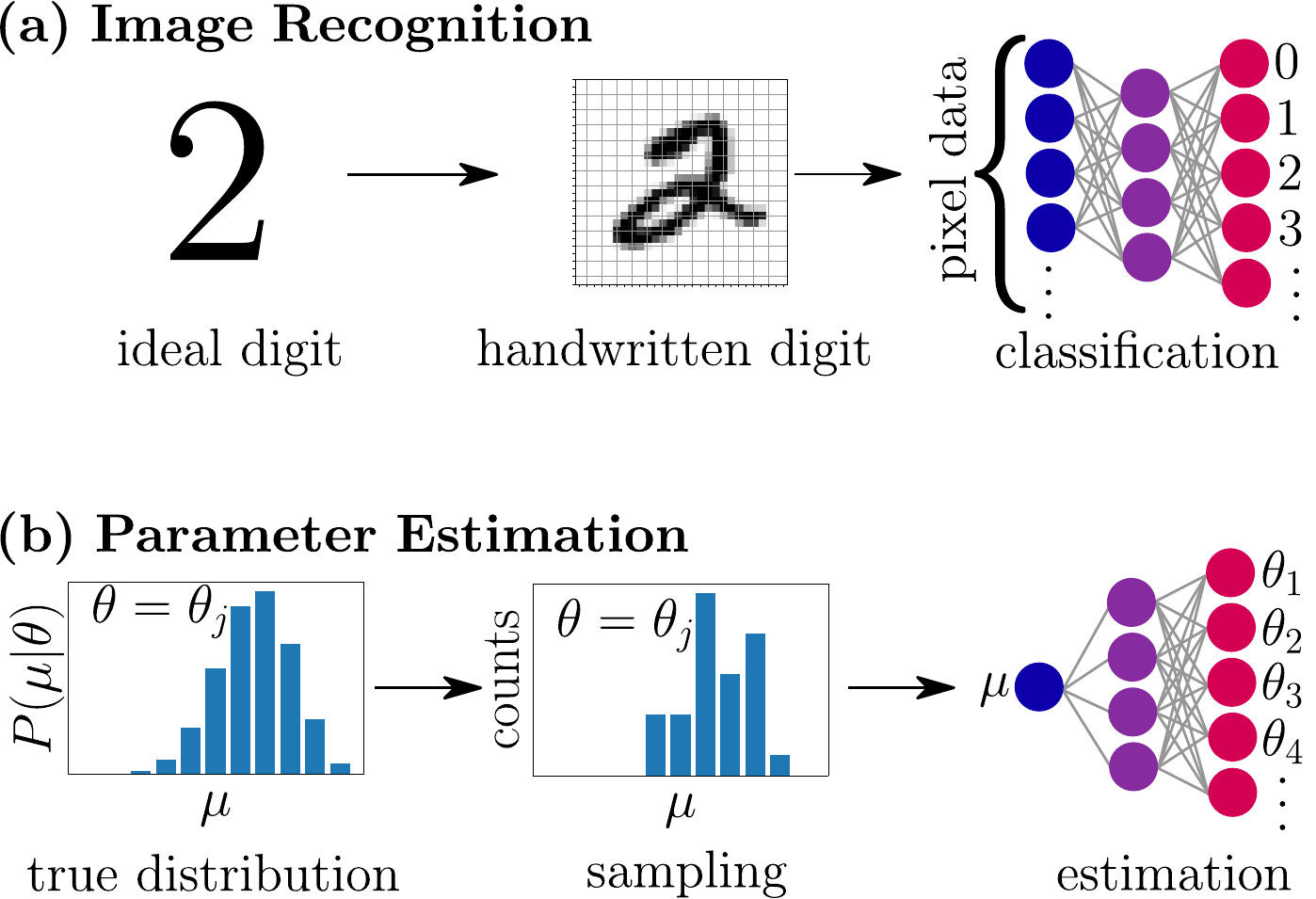}
\caption{
{\bf Parameter estimation as a classification task.}
\textbf{(a)} 
By learning the characteristic features of ideal digits directly from training examples \cite{lecun2010mnist}, the network can correctly classify handwritten digits with high accuracy.
The network provides the conditional probability $P({\rm digit}|{\rm data})$ that the image is assigned to a certain ideal digit (2 in this example) given the input pixel data.
\textbf{(b)} 
In parameter estimation, the network input is the result $\mu$ of a measurement made at the output of a quantum sensor. In analogy with the digits 0-9, the true (but unknown) probability distribution $P(\mu\vert \theta_j)$ represents a category, labelled by the discrete parameter $\theta_j$. A handwritten digit is analogous to a crude sampling from this distribution, used to train the network. Then, the output layer would assign a conditional probability $P(\theta_j \vert \mu)$ that a particular classification is correct, given the observed result $\mu$.}
\label{Figure1}
\end{figure}

%%%%%%%%%%%%%%%%%%%%%%%%%%%%%%%%%%%%%%%%%%%%%%%%%%%%%%%%%%%%%

In this manuscript we provide a machine-learning approach to BPE.
We propose that parameter estimation can be formulated as a classification task -- similar to the identification of handwritten digits, see Fig.~\ref{Figure1} --
able to be performed efficiently with supervised learning techniques based on artificial neural networks~\cite{NielsenBOOK, MurphyBOOK, MethaPR2019}.
Classification problems are naturally Bayesian: for instance, the output of the classification network in Fig.~\ref{Figure1}(a) is the probability that the handwriting is one of the digits $0, ..., 9$, in this case a well-trained network should assign the highest probability to the digit 2. Analogously, we design a neural network adapted for parameter estimation whose output is, naturally, a Bayesian parameter distribution.
Based on this interpretation, we provide a theoretical framework that enables a network to be trained using the outcome of individual measurement results.
This training provides set of Bayesian distributions for each possible experimental outcome and a Bayesian prior that we unambiguously identify and directly link to the training of the network. 
These Bayesian distributions and prior are subsequently multiplied, depending on experimental outcomes, and used to perform BPE for the estimation of an arbitrary unknown parameter.
We show that our BPE protocol is asymptotically unbiased and consistent: it obeys relevant Bayesian bounds~\cite{LiENTROPY2018} dictated, in our examples, by quantum and statistical noise.
Our method is tested on a variety of quantum states, demonstrating that classical sensitivity limits can be surpassed when using entangled states.
Crucially, the neural network needs to be trained with a relatively small 
amount of data and thus provides a practical 
advantage over the standard calibration-based BPE.

Although there is a significant body of literature on the application of machine learning techniques to solve problems in quantum science \cite{DunjkoRPP2018, CarleoRMP2019}, quantum sensing has received relatively little attention~\cite{PolinoARXIV}. 
Current studies have mainly focused on the optimization of adaptive estimation protocols~\cite{HentschelPRL2010, HentschelPRL2011, LovettPRL2013, LuminoPRAPP2018, XiaoSR2019, XuNPJ2019, PalittapongarnpimPRA2019, PengPRA2020, SchuffNJP2020, FidererARXIV2020}, improved readout for magnetometry \cite{SantagatiPRX2019, QianAppPhysLett2021} and state preparation \cite{HainePRL2012}. Similar tasks such as tomography \cite{GrossPRL2010, XuARXIV2018, TorlaiNATPHYS2018, QuekARXIV, XinNPJ2019, CarrasquillaNATMI2019, MacaronePalmieriNPJ2020}, learning quantum states \cite{SpagnoloSCIREP2017, RocchettoSCIADV2019, Yu2019, Aaronson2019, TorlaiARXIV2019, FlurinPRX2020}, Hamiltonian estimation \cite{GranadeNJP2012, WangNATPHYS2017, WangARXIV2019, WozniakowskiARXIV2020} and state discrimination \cite{YouARXIV2019, GebhartPRR2020} have also been considered.
Neural networks have been applied in the context of parameter estimation with the aim to infer/forecast noisy signals \cite{GreplovaARXIV2017, LiuJPHYSB2019, KhanahmadiARXIV}, and for the calibration of a frequentist estimator directly from training data \cite{CiminiPRL2019}.  
Unlike these approaches, we show here that a properly trained neural network naturally performs 
BPE without any assumptions about the system.
The machine-learning-based parameter estimation illustrated in this manuscript can be readily applied for data analysis in current quantum sensors, providing all the important advantages of BPE, while enjoying less stringent calibration/training requirements.
The method applies to any (mixed or pure) state and measurement observable.
In practical applications, noise and decoherence that affect the apparatus are directly included (via the training process) in the Bayesian posterior distributions which therefore fully account for experimental imperfections.

\section{Results}

%%%%%%%%%%%%%%%%%%%%%%%%%%%%%%%%%%%%%%%%%%%%%%%%%%%%%%%%%%%%%%%%%%%%
%%.. figure 2
%%%%%%%%%%%%%%%%%%%%%%%%%%%%%%%%%%%%%%%%%%%%%%%%%%%%%%%%%%%%%%%%%%%%%
\begin{figure*}[t!] 
\centering
\includegraphics[width=\textwidth]{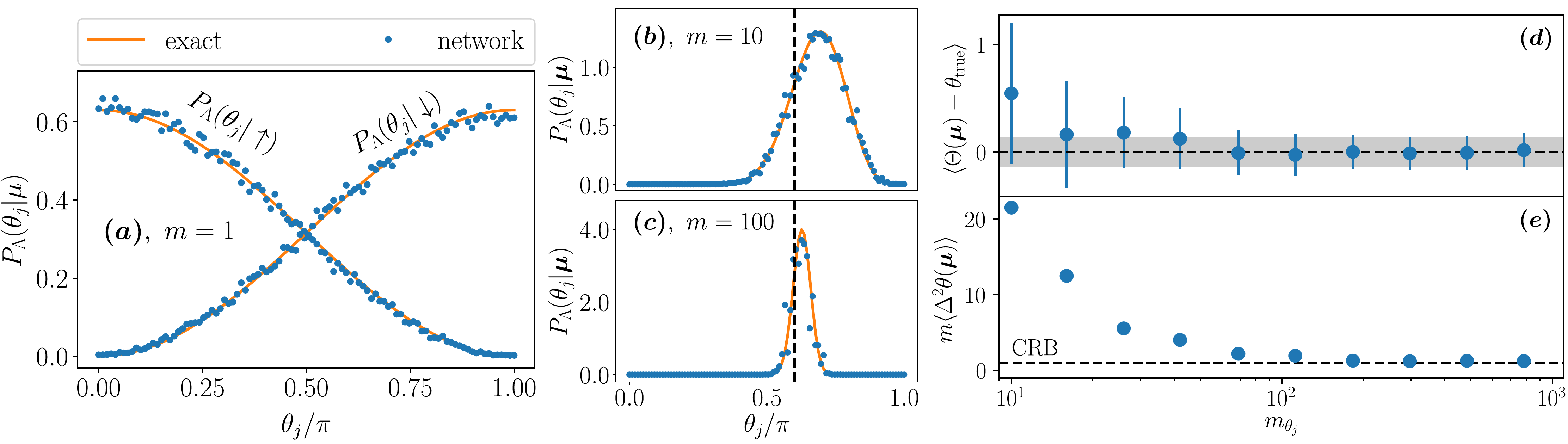}
\caption{{\bf Bayesian inference performed with a neural network.} 
Here we show results of BPE for the pedagogical example of a single qubit (see Methods and text). 
The output layer $P_\Lambda(\theta_j \vert \mu)$ of a uniformly-trained network and finite training data (blue dots, details in Methods) 
compared to the 
exact Bayesian distribution $P(\theta\vert \mu) = P(\mu\vert \theta)P(\theta)/P(\mu)$ (orange line), 
where $P(\theta) = 1/\pi$ and $P(\mu)$ provides normalization.
{\bf (a)} Bayesian posterior probabilities corresponding to the single measurement event $\uparrow$ and $\downarrow$.
Panels {\bf (b)} and {\bf (c)} show the Bayesian posterior distributions Eq.~(\ref{Ppost1}) for $m=10$ and $m=100$ repeated measurement events respectively. These are obtained from the $m=1$ posterior distribution with Eq.~(\ref{Ppost1}). 
We set $\theta_{\rm true}=0.6\pi$ (black dashed vertical lines) and randomly generate a sequence of results:
 $\boldsymbol{\mu} = \{m_{\uparrow, \theta_{\rm true}}, m_{\downarrow, \theta_{\rm true}} \}=\{ 3,7 \}$ in $\boldsymbol{(b)}$, $\boldsymbol{\mu} = \{ 29, 71 \}$  in $\boldsymbol{(c)}$.
$\boldsymbol{(d)}$ Mean value of the maximum a-posterior estimators
as a function of the training size $m_{\theta_j}$.
The shaded region is the CRB (here $\Delta^2 \theta_{\rm CRB} = 1/m$), and the error bars are the mean posterior variance, shown explicitly in $\boldsymbol{(e)}$.
In {\bf (d,e)} we fixed $m=50$.}
\label{Figure2}
\end{figure*}

%%%%%%%%%%%%%%%%%%%%%%%%%%%%%%%%%%%%%%%%%%%%%%%%%%%%%%%%%%%%%%
%%%%%%%%%%%%%%%%%%%%%%%%%%%%%%%%%%%%%%%%%%%%%%%%%%%%%%%%%%%%%%

In a general parameter estimation problem, a probe state $\rho$ undergoes a 
transformation that depends on an unknown parameter $\theta_{\rm true}$.
The goal is to estimate $\theta_{\rm true}$ from measurements performed on the output state $\hat{\rho}_{\theta_{\rm true}}$.
A detection event $\mu$ occurs with probability $P(\mu \vert \theta_{\rm true})= {\rm Tr}[\hat{\rho}_{\theta_{\rm true}} \hat{E}_{\mu}]$, where  
$\{ \hat{E}_{\mu} \}$ is a complete set of positive, $\hat{E}_\mu \geq 0$, and complete, $\sum_\mu \hat{E}_\mu  =1$ operators ~\cite{BraunsteinPRL1994}. %,NielsenChuangBOOK}.

The parameter estimation discussed in this manuscript is divided in two parts: i) a neural network is trained and ii) Bayesian estimation performed on a test set, which we detail below. A test set refers to an arbitrary sequence of measurement results $\boldsymbol{\mu}$ of length $m$, possibly different to the number of measurements found in the training set.
To build intuition we first illustrate the theory with a pedagogical example consisting in the estimation of the rotation angle of a single qubit state $\exp \left( -i \sigma_y \theta/2 \right) | \uparrow \rangle$, ($\sigma_{x,y,z}$ are the usual Pauli matrices and $| \uparrow \rangle$, $\ket{\downarrow}$ are eigenstates of $\sigma_z$). The rotation angle $\theta$ is estimated by projecting the output state $|\psi(\theta) \rangle = \exp \left( -i \sigma_y \theta/2 \right)| \uparrow \rangle$
on $\hat{\sigma}_z$.
The two possible output results, $\mu = \uparrow, \downarrow$, can occur with probability 
$P(\uparrow|\theta) = \cos^2(\theta/2)$ and $P(\downarrow|\theta) = 1-P(\uparrow|\theta)$, respectively, which are monotonic over the interval $\theta_j \in [0, \pi]$.Aside of being purely pedagogical, such a system is relevant to NV center magnetometers \cite{HincksNJP2012, BonatoNATNANO2015, SantagatiPRX2019, AhronSR2019, SchwartzSR2019}
Later, we generalise to systems of many qubits, in separable and entangled states, eventually including noise during state preparation and/or in the output measurement.

\subsection{Training of the neural network} 

First, the parameter domain is discretized to form a uniform grid of $d$ points $\theta_1, ..., \theta_d$ which are assumed to be perfectly known. The training set consists of $m_{\theta_j}$ measurements performed at each $\theta_j$. For example, the training set for a single qubit would contain $d$ tuples $\{ m_{\uparrow, \theta}, m_{\downarrow, \theta}\}$, where $m_{\mu, \theta}$ 
is the number of times the result $\mu= \uparrow, \downarrow$ was observed at a particular $\theta$. During training, the network is shown all $m_{\rm train}=\sum_{j=1}^d m_{\theta_j}$ measurement results $\mu$, along with the labels $\theta_j$ that are sampled from the (unknown) joint distribution~\cite{DunjkoRPP2018}, 
\begin{equation} \label{eq:joint}
P(\mu, \theta_j) = P(\mu \vert \theta_j) P(\theta_j).
\end{equation}
Here, $P(\mu \vert \theta_j)$ is the probability to observe a measurement result $\mu$ when the parameter is set to $\theta_j$. This distribution fully characterizes the experimental apparatus (including all sources of noise and decoherence), is typically unknown to the experimentalist and is never seen by the network. Additionally, the probabilities $P(\mu \vert \theta_j)$ need not be sampled uniformly in $\theta_j$, which may also have some distribution $P(\theta_j)$.

Via the optimization of weights and links of artificial neurons,
the network attempts to learn
the conditional probability 
$P_{\Lambda}(\theta_j \vert \mu)$ that gives the degree of certainty that $\theta_j$ is the correct label given the particular $\mu$ shown during training. 
This is the essential idea of supervised learning.
Here, the subscript $\Lambda$ denotes the dependence of the output on the randomly chosen initial network, the training algorithm, and the training data itself. 
In Fig.~\ref{Figure2}(a) we show the two possible outputs of the network for the single qubit example: that is $P_\Lambda(\theta_j \vert \uparrow)$ and 
$P_\Lambda(\theta_j \vert \downarrow)$ (blue dots), as a function of the label set $\theta_1, ..., \theta_d$ in $[0, \pi]$.

\subsection{Bayesian inversion and prior distribution}

Here, we recognize that the output of the neural network, $P_{\Lambda}(\theta_j \vert \mu)\delta \theta$, can be interpreted as a Bayesian posterior distribution. As we have discretised the continuous random variable $\theta$, it is necessary to account for the grid spacing
$\delta \theta = \theta_d/(d-1)$.
We show that the posterior distribution is formally obtained from the Bayes rule,
\begin{equation} \label{eq:truebayes}
P_{\Lambda}(\theta_j \vert \mu) = \frac{P_{\Lambda}(\mu \vert \theta_j) P_{\Lambda}(\theta_j)}{P_{\Lambda}(\mu)}.
\end{equation}
 We emphasise that the Bayesian inversion in Eq.~(
 \ref{eq:truebayes}) is performed indirectly by the network, which does not have access to any of the quantities on the right-hand side of Eq.~(\ref{eq:truebayes}). 
$P_{\Lambda}(\mu)$ normalises the posterior distribution,  $\sum_{j=1}^{d} P_\Lambda(\theta_j \vert \vect{\mu})  \delta \theta =1$ and 
$P_\Lambda(\theta_j)$ is called the prior
which plays a conceptual, as well as a practical role. Throughout this manuscript we are treating possible measurement results $\mu$ as a discrete random variable.  

We calculate $P_\Lambda(\theta_j)$ from its definition as the marginal distribution, $P_\Lambda(\theta_j) = \sum_{\mu} P_\Lambda(\theta_j \vert \mu) P_\Lambda(\mu)$ with the sum extending over all possible 
measurement results $\mu$. As $P_\Lambda(\mu)$ is also unknown, we can eliminate it by again inserting the marginal expression $P_\Lambda(\mu) = \sum_{k=1}^{d-1} P_\Lambda(\mu \vert \theta_k) P_\Lambda(\theta_k) \delta \theta$, which results in the implicit integral equation
\beq \label{Ptheta}
P_\Lambda(\theta_j) = \sum_{\mu} P_\Lambda(\theta_j \vert \mu) \sum_{k=1}^{d} P_\Lambda(\mu \vert \theta_k) P_\Lambda(\theta_k) \delta \theta
\eeq
Equation (\ref{Ptheta}) is a consistency relation that can be solved for $P_\Lambda(\theta_j)$, given the network output $P_\Lambda(\theta_j \vert \mu)$ and the likelihood function $P_\Lambda(\mu \vert \theta_j)$.
The relation Eq.~(\ref{Ptheta}) can be solved for $P_\Lambda(\theta_j) \equiv p_j$ by recasting it as an eigenvalue problem $\vect{A} \vect{p} = \vect{0}$, for the matrix
\be \label{EigEq}
\vect{A}_{jk} = \delta_{jk} - \sum_{\mu} P_\Lambda( \theta_j \vert \mu) P_\Lambda(\mu \vert \theta_k) \delta \theta,
\ee 
where $\delta_{jk}$ is the Kronecker delta. 
To evaluate Eq.~(\ref{EigEq}) the likelihood $P_\Lambda(\mu \vert \theta_k)$ is needed, however the network only provides $P_\Lambda( \theta_j \vert \mu)$.
For a sufficiently well trained network, we can approximate it with 
the ideal likelihood distribution, $P_\Lambda(\mu \vert \theta_k) \approx P(\mu\vert\theta_k)$, which is either known from theory, or else
can be well approximated by the relative frequencies observed in the training data $P_\Lambda(\mu \vert \theta_j) \approx m_{\mu,\theta_j}/m_{\theta_j}$.
We have found that the prior calculation in Eqs.~(\ref{Ptheta}) and (\ref{EigEq}) is robust to the choice of $P_\Lambda(\mu \vert \theta_k)$. 

\begin{figure} 
\centering
\includegraphics[width=\columnwidth]{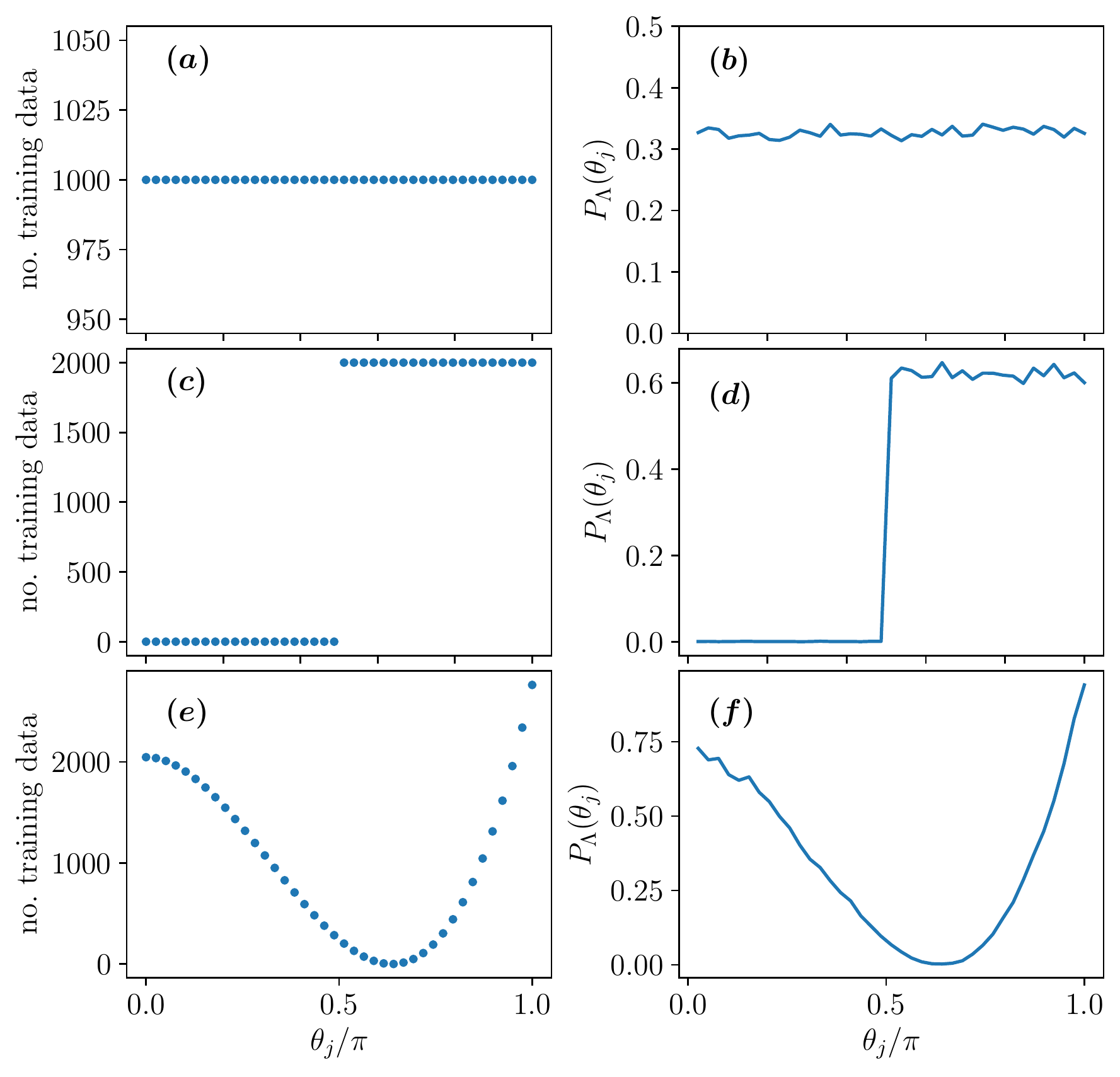}
\caption{
{\bf Prior vs. training distribution.}
In the left column panels we show examples of distribution of training data, $m_{\theta_j}$ as a function of $\theta_j$. 
The right column panels show the corresponding Bayesian prior distribution $P_\Lambda(\theta_j)$. In all three examples the total number of measurements are held fixed. Specifically, in $\boldsymbol{(a,b)}$ $m_{\theta_j}$ are distributed uniformly, resulting in a flat prior. In $\boldsymbol{(c,d)}$ data is distributed according to a step function, clearly resulting in a prior that is zero-valued over the part of the domain where no measurements were performed. Finally, in $\boldsymbol{(e,f)}$ a smooth distribution $q(\theta_j)$ is used, which is clearly reproduced by the network.  }
\label{Figure3}
\end{figure}

As shown in Fig.~\ref{Figure3}, the prior $P_\Lambda(\theta_j)$ is determined by the sampling of the training data. For instance, if the training data is distributed uniformly ($m_{\theta_j} = m$ independent of $\theta$),
then $P_\Lambda(\theta_j)$ is flat, as in Fig. \ref{Figure3} $\boldsymbol{(a,b)}$. 
A non-flat prior could be achieved by choosing a non-uniform distribution of training measurements. For instance, if $m_{\rm train}$ is the total number of measurements collected in the training set, the number of measurements $m_{\theta_j}$ at each $\theta_j$ could be distributed according to $m_{\theta_j} = m_{\rm train} q(\theta_j)$ where $q(\theta_j)$ is a 
positive function of $\theta_j$ with $\sum_{j=1}^d q(\theta_j)=1$. In this case, a well-trained network will {\it learn} a prior well approximated by $P_\Lambda(\theta_j) \approx q(\theta_j) $. Two examples are shown in Fig. \ref{Figure3}, panels $\boldsymbol{(c,d)}$ and $\boldsymbol{(e,f)}$.
The grid itself could also be varied, resulting in a non-uniform grid spacing $\delta \theta_j = \theta_{j+1}-\theta_j$, which would also result in a non-flat prior. However, this is equivalent to a choice of $q(\theta_j)$ on a uniform grid. This is clearly illustrated by the step function example [Fig.~\ref{Figure3} $\boldsymbol{(c,d)}$]. Rather than $q(\theta_j)$ itself being a step function, the same result could be achieved using a flat $q(\theta_j)$ over a grid spanning $[\pi/2, \pi]$ (rather than $[0, \pi]$) but sampled at twice the density. For this reason, we consider only uniform grid spacing throughout this manuscript.
The prior thus retains the subjective nature that characterizes the Bayesian formalism: here, this subjectivity is associated with the arbitrariness in the collection of the training data.

%%%%%%%%%%%%%%%%%%%%%%%%%%%%%%%%%%%%%%%%%%%%%%%%%%%%%%%%%%%%%%%%%%%%%%%%%%
%%.. figure 3
%%%%%%%%%%%%%%%%%%%%%%%%%%%%%%%%%%%%%%%%%%%%%%%%%%%%%%%%%%%%%%%%%%%%%%%%%%
%%%%%%%%%%%%%%%%%%%%%%%%%%%%%%%%%%%%%%%%%%%%%%%%%%%%%%%%%%%%%%%%%%%%%%%%%%

\subsection{ Network-based BPE}

The training of the network gives access to the single-measurement ($m=1$) conditional probabilities $P_\Lambda(\theta_j \vert \mu)$ and 
the prior distribution $P_\Lambda(\theta_j)$.
We thus proceed with the estimation of an unknown parameter $\theta_{\rm true}$
(of course in the numerical experiment $\theta_{\rm true}$ is known but this information is never used).
Notice that $\theta_{\rm true}$ does not need to coincide with one of the grid values $\theta_j$. We sample $m$ random measurement results $\vect{\mu} = \mu_1, ..., \mu_m $ from $P(\mu \vert \theta_{\rm true})$. 
The Bayesian posterior distribution corresponding to the sequence $\boldsymbol{\mu}$ is 
\be \label{Ppost1}
P_\Lambda(\theta_j \vert \vect{\mu}) = \mathcal{N}  P_\Lambda(\theta_j) \prod_{i=1}^m \tilde{P}_\Lambda(\theta_j \vert \mu_i),
\ee 
where $\tilde{P}_\Lambda(\theta_j \vert \mu_i) = P_\Lambda(\theta_j \vert \mu_i)/P_\Lambda(\theta_j)$ and $\mathcal{N}$ is a normalisation factor (obtained numerically). 
For concreteness, in the single-qubit example, if a sequence of $m$ measurements gives $m_\uparrow$ results $\uparrow$ and  $m_\downarrow = m - m_\uparrow$ results $\downarrow$,
the corresponding Bayesian probability distribution is 
$P_\Lambda(\theta_j \vert \vect{\mu}) = \mathcal{N} P_\Lambda(\theta_j) \tilde{P}_\Lambda(\theta_j \vert \downarrow)^{m_\downarrow} \tilde{P}_\Lambda(\theta_j \vert \uparrow)^{m_\uparrow} $, see Fig.~\ref{Figure2} $\boldsymbol{(b,c)}$.
Equation (\ref{Ppost1}) represents an update of knowledge about $\theta_{\rm true}$ as measurements are collected.
Such Bayesian update is based on single-measurement distributions $P_\Lambda(\theta_j \vert \mu)$ and the prior $P_\Lambda(\theta_j)$.
Indeed, a key advantage of our method is that, while the network is trained with single ($m=1$) measurement events, 
the Bayesian analysis can be performed, according to Eq.~(\ref{Ppost1}), for arbitrary large $m$.
In other words, we do not need to train the network for each $m$: the network is 
trained for $m=1$, which guarantees the optimal use of training data.
We emphasize that the prior $P_\Lambda(\theta_j)$ in Eq.~(\ref{Ppost1})
is obtained by solving Eq.~\eqref{Ptheta}:
even for a uniform training, the Eq.~\eqref{Ptheta} gives a better results  compared to $P(\theta_j)=1/\pi$.

Given $P_\Lambda(\theta_j \vert \vect{\mu})$ we can estimate $\theta_{\rm true}$ by, for instance,
\be \label{BayesianEst}
\Theta(\boldsymbol{\mu}) = {\rm arg} \Big[ \max_\theta P_\Lambda(\theta_j | \boldsymbol{\mu}) \Big],
\ee
where the corresponding parameter uncertainty is quantified by the posterior variance
\begin{equation} \label{PosteriorVar}
    \Delta^2 \theta(\boldsymbol{\mu}) = \sum_{j=1}^{d} P_\Lambda(\theta_j \vert \vect{\mu}) \left[\Theta(\boldsymbol{\mu}) - \theta_j \right]^2 \delta \theta,
\end{equation}
which assigns a confidence interval to any
measurement sequence $\boldsymbol{\mu}$.
In a sufficiently well trained network, as the number of measurements $m$ increases, $P_\Lambda(\theta_j \vert \boldsymbol{\mu})$ converges to the 
Gaussian distribution \cite{LehmannBOOK, PezzeBOOK2014}
\be \label{Pasympt}
P_\Lambda(\theta_j \vert \vect{\mu}) \approx \sqrt{\frac{m F(\theta_{\rm true})}{2 \pi }} e^{-m F(\theta_{\rm true}) (\theta_j -\theta_{\rm true})^2/2},
\ee
centered at the true value $\theta_{\rm true}$ and with variance $1/m F(\theta_{\rm true})$, where 
\be
F(\theta) = \sum_{\mu} \frac{1}{P(\mu \vert \theta)} \bigg( \frac{d P(\mu \vert \theta)}{d \theta} \bigg)^2
\ee
is the Fisher information. $F(\theta)$ provides a frequentist bound on the precision of a generic estimator $\Delta^2 \theta \geq\Delta^2\theta_{\rm CRB} = 1/m F(\theta_{\rm true})$, called the Cram{\'e}r-Rao bound. This behavior is clearly exhibited by the network in Fig.~\ref{Figure2} $\boldsymbol{(b,c)}$: the distribution narrows as a function of $m$ and centres around $\theta_{\rm true}$.
The result Eq.~(\ref{Pasympt}) is valid for a sufficiently dense grid (i.e. $\delta \theta \ll 1/\sqrt{m F(\theta_{\rm true})}$) and in an appropriate phase interval around $\theta_{\rm true}$,
and holds for any prior distribution $P(\theta_j)$, provided that $P(\theta_j)$ is non-vanishing around $\theta_{\rm true}$. 
By repeating the measurements and using Eq.~(\ref{Ppost1}), we can thus gain a factor $\sqrt{m}$ in sensitivity, $\Delta \theta \sim 1/\sqrt{m}$, without requiring either additional training data
or additional training for each $m$. 
In other words, a single network can be used to provide an estimate for any number of repeated measurements $m$, limited only by the grid size, meaningful for $\Delta \theta \gg \delta \theta$.
In the opposite limit, and thus for $m \gg F(\theta_{\rm true})/(\delta \theta)^2$, the estimation is biased, namely $\vert \langle \Theta(\mu) - \theta_{\rm true} \rangle \vert \gtrsim \sqrt{\langle \Delta^2 \theta \rangle}$. The brackets $\langle \cdots \rangle$ denote the average over the likelihood function $P(\boldsymbol{\mu} \vert \theta_{\rm true})$.
The presence of an asymptotic bias is intrinsic of Bayesian estimation on a finite grid, when $\theta_{\rm true}$ does not coincide with one of the grid points.
The effect is present also when using ideal probabilities (namely in the limit $m_{\rm train}\to \infty$) and it is not associated with the neural network.
Of course, insufficient training produces a network that poorly generalises to larger $m$. Figure~\ref{Figure2} $\boldsymbol{(d,e)}$ shows convergence to the expected asymptotic result as a function of the number of training examples $m_{\theta_j}$, for a fixed number of measurement events $m=50$.

%%%%%%%%%%%%%%%%%%%%%%%%%%%%%%%%%%%%%%%%%%%%%%%%%%%%%%%%%%%%%%%%%%%%%%%%%%
%%.. figure 4
%%%%%%%%%%%%%%%%%%%%%%%%%%%%%%%%%%%%%%%%%%%%%%%%%%%%%%%%%%%%%%%%%%%%%%%%%%
\begin{figure*}[t!] 
\centering
\includegraphics[width=\textwidth]{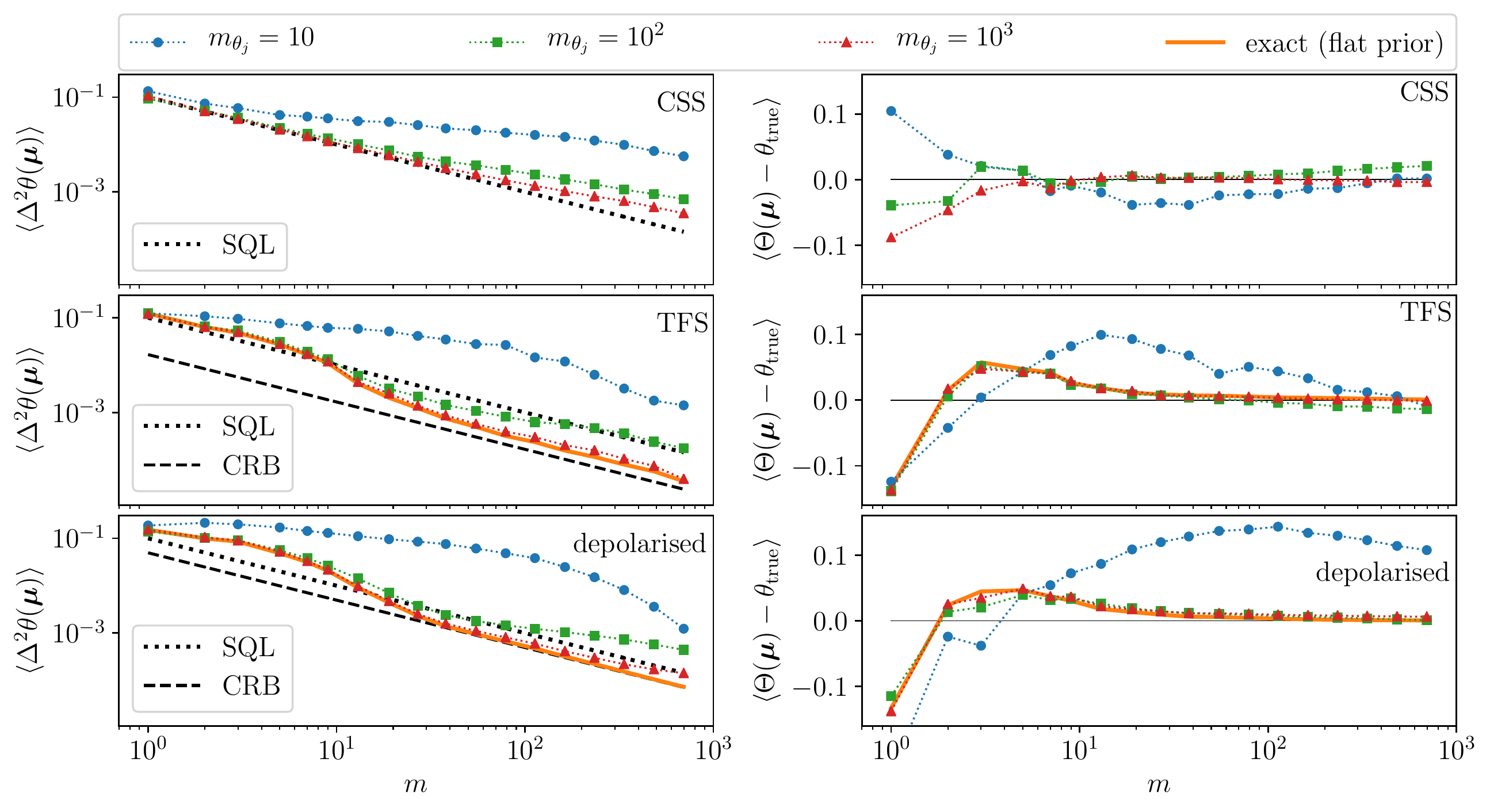}
\caption{
{\bf Consistency and efficiency for many-qubit states.}
Here we plot the mean Bayesian posterior variance 
Eq.~(\ref{PosteriorVar}) (left panels) and the 
bias $\langle \Theta(\vect{\mu}) - \theta_{\rm true} \rangle$ (right)
as a function of the number of repeated measurements $m$.
Top panels consider a CSS, the middle panels a TFS, and the bottom panels a depolarised TFS, all with $N=10$ qubits.
For all states, three networks are trained with 
uniform data $m_{\theta_j}=10$ (blue dots), 
$m_{\theta_j}=10^2$ (green squares),
$m_{\theta_j}=10^3$ (red triangles).
The solid orange lines are 
exact result, obtained from Bayes rule using the true probabilities $P(\mu \vert \theta)$ and a flat prior.
Dashed black lines are the standard quantum limit (SQL) and the frequentist Cram{\'e}r-Rao bounds (CRB).
Here $\theta_{\rm true}=0.3\pi$ (not found in the training grid) and all results are averaged over $10^3$ randomly generated measurement sequences of length $m$. 
See Methods for details on the network parameters.}
\label{Figure4}
\end{figure*}

%%%%%%%%%%%%%%%%%%%%%%%%%%%%%%%%%%%%%%%%%%%%%%%%%%%%%%%%%%%%%%%%%%%%%%%%%%

%%%%%%%%%%%%%%%%%%%%%%%%%%%%%%%%%%%%%%%%%%%%%%%%%%%%%%%%%%%%%%%%%%%%%%%%%%
%%.. figure 5
%%%%%%%%%%%%%%%%%%%%%%%%%%%%%%%%%%%%%%%%%%%%%%%%%%%%%%%%%%%%%%%%%%%%%%%%%%
\begin{figure} [h!]
\centering
\includegraphics[width=\columnwidth]{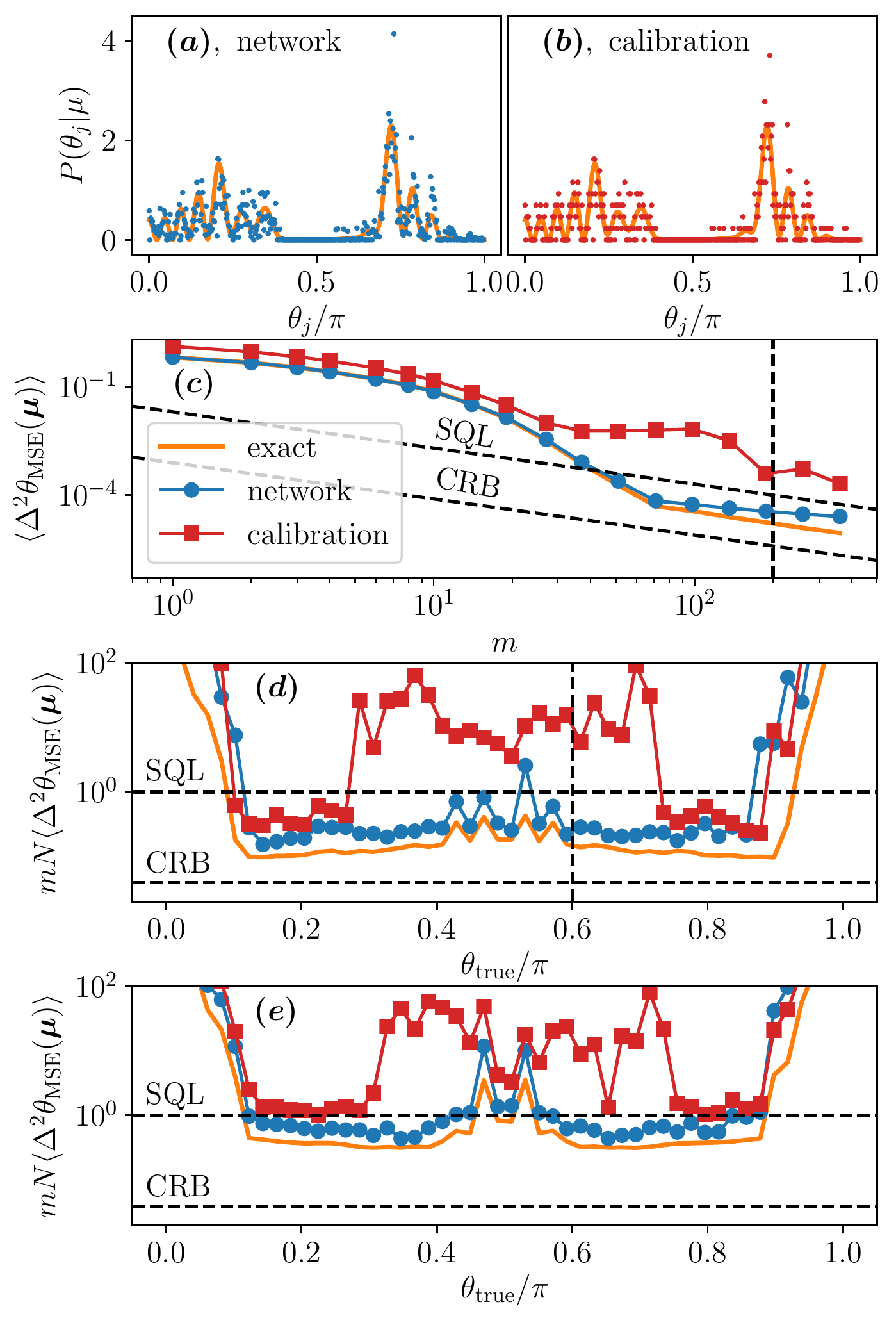}
\caption{{\bf Comparison with calibration-based BPE}. Comparison between neural network-based BPE to calibration-based BPE using the same number of training/calibration measurements for an over-squeezed state of $N=50$ qubits, as discussed in the main text. {\bf (a,b)} Example of a single-shot posterior for $\mu=15$, learned directly by the network {\bf (a)} or inferred from the training data  (assuming a flat prior) {\bf (b)}, both from the same set of $m_{\theta_j}=100$ training/calibration measurements at each phase. {\bf (c,d,e)} The posterior MSE shows the advantage of the neural network procedure over the calibration, for $m_{\theta_j}=500$. In {\bf (c)} the network is shown to stay closer to the true posterior MSE (solid orange) over a much larger range of $m$ values than the calibration, at a fixed value of the true phase $\theta_{\rm true}=0.6\pi$ [not an element of the training grid, vertical black dashed line in {\bf (d)}]. In {\bf (d,e)} we study the performance on a grid of $\theta_{\rm true}$ values spanning the entire estimation domain $[0,\pi]$ that were not found in the training grid. In {\bf (d)} the advantage is found to persist over many values of $\theta_{\rm true}$, at $m=200$ shots [vertical black dashed line in {\bf (c)}]. Finally {\bf (e)} includes the effects of finite detection resolution $\Delta \mu^2=0.25$, but otherwise parameters are the same as in {\bf (d)}. The likelihood average is approximated by averaging over $10^4$ randomly chosen measurement sequences $\boldsymbol{\mu}$. See Methods for details on numerical parameters.}
\label{Figure5}
\end{figure}

%%%%%%%%%%%%%%%%%%%%%%%%%%%%%%%%%%%%%%%%%%%%%%%%%%%%%%%%%%%%%%%%%%%%%%%%%%

The strategy of classifying a sequence $\boldsymbol{\mu}$ following training based on single measurement results $\mu$ only ($\mu=\uparrow,\downarrow$ for the single-qubit case), is a key difference between this work and typical supervised learning problems such as image recognition \cite{NielsenBOOK, MurphyBOOK, MethaPR2019}.
With image recognition there is a risk that during training a network will merely memorize the training images, and poorly generalise to unseen images (this is called overfitting). The single-measurement training that we use avoids this problem.  
Instead, our network is expected to generalise from the single measurement results seen during training, to sequences with $m>1$ via Eq.~(\ref{Ppost1}). Therefore, the network will never be asked to perform a prediction on an input $\mu$ not found in the training set (which will also only ever contain e.g. $\mu=\uparrow,\downarrow$, as in the single-qubit example). Rather, if the machine-learned Bayesian posterior for the single-measurements $\mu$ is noisy or imperfect, this error will quickly compound when Eq.~(\ref{Ppost1}) is applied.
Therefore, it is important to compute metrics relevant to parameter estimation such as the mean bias or posterior variance (as in Fig.~\ref{Figure4}). 

\subsection{Application to many-qubit states}

In this section we extend our procedure to systems of $N$ qubits and demonstrate its effectiveness for both separable and entangled states. 
We introduce the collective spin operators $\hat{J}_k=\sum_{i=1}^N \sigma_k^{(i)}/2$, where $\sigma_k^{(i)}$ is the $k$th Pauli matrix for the $i$th qubit. Making use of these observables, the generalisation from a single qubit to many qubits is straightforward: the network is trained to recognise the result of a single $\hat{J}_z$ measurement with $N+1$ possible outcomes. The Bayesian posterior for many measurements is then obtained from Eq.~(\ref{Ppost1}). We consider phase-dependence encoded by a rotation about $\hat{J}_y$, which is equivalent to a Mach-Zehnder interferometer~\cite{PezzeRMP2018}. In Fig.~\ref{Figure4} we apply our method to a coherent-spin state (CSS) $|\rm CSS \rangle = \ket{\downarrow}^{\otimes N}$ (top panels), a twin-Fock state (TFS) 
given by the symmetrized combination of $N/2$ spin-up and $N/2$ spin-down particles $|\rm TFS \rangle = {\rm Symm}\{ \ket{\downarrow}^{\otimes N/2}, \ket{\downarrow}^{\otimes N/2} \}$
(middle panels) and a depolarised TFS $\hat{\rho} = (1-\epsilon)|\rm TFS \rangle \langle \rm TFS | + \epsilon I/(N+1)$ where $I$ is the identity matrix (in the subspace of permutation-symmetric states) and $\epsilon=0.1$ (bottom panels).
We quantify the performance of the network by the mean posterior variance $\langle \Delta^2\theta(\boldsymbol{\mu}) \rangle$ and bias $\langle \Theta(\boldsymbol{\mu})-\theta_{\rm true} \rangle$, averaged over all possible measurement sequences $\boldsymbol{\mu}$.
For all three states, Fig.~\ref{Figure4} shows that our neural network-based BPE is asymptotically efficient and unbiased when tested on a $\theta$ not found in the training grid. As expected for the CSS, the posterior variance saturates the standard quantum limit on average (SQL, $\Delta^2\theta_{\rm SQL} = 1/mN$). Similarly, the TFS posterior variance (\ref{PosteriorVar}) overcomes the SQL and approaches, on average, the Cram\'er-Rao bound $\Delta^2 \theta_{\rm CRB} = 1/[m N (N/2+1)]$ in the limit of many repeated measurements $m$. The same is true for the depolarised TFS, demonstrating that our neural network-based BPE is also applicable to mixed states.
Furthermore, on average, the estimator (\ref{BayesianEst}) gives the true value of the parameter,
as expected -- so long as the training set is sufficiently large relative to the desired number of measurements $m$.
In particular, networks that are shown more measurements during training are better able to generalise to large $m$.

\subsection{Comparison to calibration-based BPE}

It is natural to ask how well the network compares to conventional (calibration-based) BPE~\cite{PezzePRL2007, KrischekPRL2011, XiangNATPHOT2011} making use of the same training data. Consider a training set where $m_{\theta_j}$ measurements are performed at each $\theta_j$, with result $\mu$ occurring $m_\mu$ times at this $\theta_j$. We assume a uniform distribution $m_{\theta_j}$, corresponding to a flat prior. The standard approach to either Bayesian or maximum likelihood estimation is to take this data set, and estimate the likelihood functions $P(\mu \vert \theta_j)$ using the relative frequencies $P(\mu \vert \theta_j) \approx m_{\mu,\theta_j}/m_{\theta_j} \equiv f_{\mu, \theta}$, usually aided by some kind of fitting procedure~\cite{PezzePRL2007, KrischekPRL2011}. The posterior distribution $P(\theta_j \vert \mu) $ is then obtained by choosing a prior $P(\theta_j)$ and applying Bayes theorem $P(\theta_j \vert \vect{\mu}) = P(\theta_j)\prod_{i=1}^m P(\mu_i \vert \theta_j) /P(\vect{\mu})$, where $P(\vect{\mu})$ provides normalization and $\vect{\mu} = \mu_1, ..., \mu_m$ is a measurement sequence. We call this a calibration-based Bayesian analysis.
A drawback is that it generally requires collecting a large calibration data set, such that relative frequencies 
$f_{\mu, \theta}$ well approximate the corresponding probabilities. 
A further problem is that it is not possible to associate a Bayesian probability to (rare) detection events that did not appear during the calibration, unless the probability is inferred through an arbitrary fit or interpolation procedure.
Both issues are overcome by our neural network-based BPE. 

In Figure \ref{Figure5} we compare our network-based BPE to the calibration-based BPE. We consider a multi-partite entangled, non-Gaussian state (ENGS) of $N=50$ qubits. Entanglement is generated using the one-axis twisting Hamiltonian $H_{\rm OAT} = \hbar \chi \hat{J}_z^2$ \cite{KitagawaPRA1993}, for $\chi t = 0.3\pi$ which is in the over-squeezed regime~\cite{PezzePRL2009}. Being highly non-Gaussian, it is difficult to aid the calibration with parametric curve fitting. The network on the other hand, is well suited to learning arbitrary probability distributions. Figure~\ref{Figure5} {\bf (a)} shows a typical example of a single-shot posterior distribution learned by the network, compared to the relative frequencies in Fig.~\ref{Figure5} {\bf (b)}. 
The relative frequencies are intrinsically coarse grained, e.g. in Fig.~\ref{Figure5} {\bf (b)} the resolution limit $1/m_{\theta_j}$ is visible, unlike the network which is smooth. 
In Fig.~\ref{Figure5} {\bf (c,d)} we compare
the statistically-averaged posterior mean-square error (MSE), 
\begin{equation}
\Delta^2\theta_{\rm MSE}(\boldsymbol{\mu}) = \sum_{j=1}^{d} P(\theta_j \vert \boldsymbol{\mu}) \left(\theta_{\rm true} - \theta_j \right)^2 \delta \theta,
\end{equation}
which quantifies the fluctuations in the deviation of the Bayesian estimate from $\theta_{\rm true}$
(see \cite{LiENTROPY2018} and Refs. therein).
The posterior MSE is a useful figure of merit in realistic models (either a network or a calibration attempt) because imperfections due to unavoidable noise in training/calibration data can result in an individual estimate $\Theta(\boldsymbol{\mu})$ deviating from the true value $\theta_{\rm true}$, even asymptotically. Calibration/training noise can result in positively or negatively biased estimates with equal frequency, which can lead to a deceptively low bias on average (this explains the low bias in Figure \ref{Figure4} when $m_{\theta_j}=10$).
 Figure~\ref{Figure5} {\bf (c,d)} clearly show that the neural network outperforms the calibration (see Methods for details), independently of the phase shift $\theta_{\rm true}$ or the number of measurements $m$. As a sanity check, we have verified that the calibration and the network agree well when the training set is large enough. The solid orange curve is the exact result (as would be produced by a perfect calibration/network). This is clear evidence that with limited training/calibration data, our machine learning approach can provide an advantage over conventional calibration techniques for states that are difficult or impossible to fit.
 Finally, in Fig.~\ref{Figure5} {\bf (e)} we include the effects of finite detection resolution $\Delta \mu$, which is a major limitation in large $N$ systems \cite{PezzeRMP2018}.
Modelling of detection noise is discussed in Methods. 
 Although the sensitivity is degraded, network-based BPE continues to outperform calibration-based BPE given equal training/calibration resources, see Methods for details.

\section{Discussion}

By reformulating parameter estimation as a classification task,
we have shown how to efficiently perform BPE using an artificial neural network
with an optimal use of calibration data.
The prior distribution -- which is  the characteristic trait of BPE -- is directly linked to the training process: the subjectivity of prior knowledge is reflected by the subjective choice of training strategy.

BPE offers important advantages, most notably the asymptotic saturation of the frequentist Cram\'er-Rao bound that holds regardless the statistical model. Indeed, we
have demonstrated that our strategy is consistent and efficient for both separable and entangled states of many qubits. 
Compared to other BPE protocols based on calibration data, our method is the most effective for
 non-Gaussian states. We found that our neural network-based BPE procedure can outperform standard calibration-based BPE protocols when the training/calibration data is limited and in the absence of an obvious or simple fitting functions. This advantage persists in the presence of finite detection resolution and for noisy probe states.
 In fact, our approach is the most valuable when the quantum sensor is a {\it black box}, namely when conditional probabilities of possible measurement results lack an simple explicit model based on a few fitting parameters. In this case our knowledge about the quantum sensor operations is limited to calibration data.
 
Our neural network-based BPE is readily applicable to current optical and atomic experiments, and therefore could enable BPE with entangled non-Gaussian states in current high precision quantum sensors. 
Although we focus on single-parameter estimation, our result could also be extended to the simultaneous estimation of multiple parameters. \\

{\bf Acknowledgments.}
We would like to thank V. Gebhart for useful discussions. We acknowledge funding from the project EMPIR-USOQS, EMPIR projects are co-funded by the European Unions Horizon2020 research and innovation programme and the EMPIR Participating States. 
We also acknowledge financial support from the European Union's Horizon 2020 research and innovation programme - Qombs Project, FET Flagship on Quantum Technologies grant no. 820419, and from the H2020 QuantERA ERA-NET Cofund in Quantum Technologies projects QCLOCKS and CEBBEC.

\section{Methods}

{\bf Machine learning methods.}
Throughout this manuscript we employ densely connected, feed-forward neural networks. The networks are implemented and trained using the python-based, open source package \textit{Keras} \cite{chollet2015keras}. All hidden layers use ReLU neurons (rectified linear unit). All networks have a single input neuron, which accepts a single, real number $\mu$. The number of hidden layers depends on the system, but for a single qubit a single layer of 4 neurons is sufficient (see Fig.~\ref{Figure2}). For larger and more complex states, more layers and neurons can help, as in Fig.~ \ref{Figure4} or Fig.~ \ref{Figure5}. The output layer is $d$ softmax neurons, one for each $\theta_j$ grid point, whose value is denoted $\boldsymbol{a}$, which is normalised $\sum_j a_j=1$ by construction. As we argue in the main text, the output of the network should be interpreted as a Bayesian posterior distribution,

\begin{equation}
    a_j = P_\Lambda(\theta_j \vert \boldsymbol{\mu}) \delta \theta .
\end{equation}
The training process is described in depth elsewhere, see for instance Refs. \cite{NielsenBOOK, MethaPR2019}. Briefly, the network is first initialised with random weights. For efficiency, the training set is randomly divided into subsets called mini-batches. The label $\theta_j$ is encoded as a  $d$-dimensional vector whose $k$th element is a Kronecher delta function $\delta_{jk}$. Each training element in the current mini-batch is fed into the network, and its label is used to evaluate a cost function $C$. We use the categorical cross-entropy, which for a $\boldsymbol{\mu}$ with label $\theta_j$ is simply $C=-\log \left( a_j \right)$.
$C$ is then averaged over the whole mini-batch, and minimised using the ADAM algorithm \cite{KingmaARXIV2014}. This is repeated until the entire training set is exhausted, which is called a training epoch. Typically many epochs are required to reach an optimal network.

{\bf Numerical details for Figures.}

In Fig. \ref{Figure2}, the network has a single input neuron (which takes as input the result of a single measurement $\mu$), a single hidden layer of 4 neurons and 100 output neurons (corresponding to a $\theta$ grid with 100 grid points). 
The training set contained $m_{\theta_j} = 10^3$ training measurements per grid point, evenly distributed (corresponding to a flat prior). The network was trained for 5 epochs with a mini-batch size of 128.

In Fig. \ref{Figure3},
 networks were trained to perform inference on a single qubit, and have 40 output neurons (corresponding to a $\theta$-grid of 40 points), but otherwise have the same architecture as the network in Fig. \ref{Figure2}.
Training is performed for 10 epochs with a mini-batch size of 128. The training set contains total of $m_{\rm train} = 40 \times 10^3$ measurement results. 

In Fig.~ \ref{Figure4}, the network trained for coherent-spin states had 1 input neuron, 1 hidden layer of 8 neurons, and 1000 output neurons between $0 \leq \theta_j \leq \pi$. The twin-Fock state network was more complex, 1 input neuron, 2 hidden layers with 32 neurons each, and 1000 output neurons uniformly distributed between $0 \leq \theta_j \leq \pi/2$. Training parameters are adapted to the size of the training set, which is uniform (corresponding to a flat prior). The coherent-spin state training parameters are for $m_{\theta_j} = 10,100,1000$: 60 epochs with a min-batch size of 8, 40 with 16, and 20 with 32, respectively. The twin-Fock state training parameters are for $m_{\theta_j}=10,100,1000$: 60 epochs with a min-batch size of 8, 40 with 16, and 30 with 128, respectively.

In Fig.~ \ref{Figure5}, the neural network had 3 hidden layers with 256 neurons in each, and an output grid with 2000 neurons between $0 \leq \theta_j \leq \pi$. Training was for 60 epochs with a mini-batch size of 1024. The calibration was performed by approximating the likelihood function $P(\mu \vert \theta_j)$ by the relative frequencies observed in the training data, smoothed with a cubic interpolation at twice the grid density. The interpolation was performed using \texttt{interp1d} from Python's \texttt{scipy} package. 

{\bf Finite detection resolution.}
Fig.~ \ref{Figure5} $\bold{(e)}$ also includes the effects of finite detector resolution $\Delta \mu$. Following Ref. \cite{PezzeRMP2018, PezzeBOOK2014}, detection resolution is modelled as Gaussian noise with variance $\Delta \mu^2$ and mean $\mu$. The probability of measuring the correct result $\mu$ is given detector uncertainty $\Delta \mu$ is the convolution $P(\mu \vert \theta, \Delta \mu) = \sum_{\mu'} \mathcal{C}_{\mu'} \exp\left[- (\mu-\mu')^2/2\Delta \mu^2 \right] P(\mu' \vert \theta)$ where $\mathcal{C}_{\mu'}=\left( \sum_\mu \exp\left[- (\mu-\mu')^2/2\Delta \mu^2 \right] \right)^{-1} $ normalises $P(\mu \vert \theta, \Delta \mu)$.

\section{Data Availability}

The datasets generated during and/or analysed during the current study are available from the corresponding author on reasonable request.

\section{Code Availability}

Any code used for the current study are available from the corresponding author on reasonable request.

\section{Author Contributions}

L.P. and A.S. were responsible for inception of the project, and all authors contributed to its ongoing design and development. S.P.N. wrote the code and performed the numerical analysis presented in this manuscript. All authors contributed to the writing of the manuscript.

\section{Competing Interests}

The authors declare no competing interests.


\begin{thebibliography}{0}
\expandafter\ifx\csname natexlab\endcsname\relax\def\natexlab#1{#1}\fi
\expandafter\ifx\csname bibnamefont\endcsname\relax
  \def\bibnamefont#1{#1}\fi
\expandafter\ifx\csname bibfnamefont\endcsname\relax
  \def\bibfnamefont#1{#1}\fi
\expandafter\ifx\csname citenamefont\endcsname\relax
  \def\citenamefont#1{#1}\fi
\expandafter\ifx\csname url\endcsname\relax
  \def\url#1{\texttt{#1}}\fi
\expandafter\ifx\csname urlprefix\endcsname\relax\def\urlprefix{URL }\fi
\providecommand{\bibinfo}[2]{#2}
\providecommand{\eprint}[2][]{\url{#2}}

\end{thebibliography}


\begin{thebibliography}{100}

\bibitem{PezzeRMP2018}
L. Pezz\`e, A. Smerzi, M.K. Oberthaler, R. Schmied and P. Treutlein,
Quantum metrology with nonclassical states of atomic ensembles,
Rev. Mod. Phys. {\bf 90}, 035005 (2018).

\bibitem{DegenRMP2017}
C.L. Degen, F. Reinhard and P. Cappellaro, 
Quantum sensing,
Rev. Mod. Phys. {\bf 89}, 035002 (2017).

\bibitem{SchnabelPR2017}
R. Schnabel, 
Squeezed states of light and their applications in laser interferometers,
Physics Reports {\bf 684}, 1-51 (2017).

\bibitem{TsePRL2019}
M. Tse {\it et al.},
Quantum-Enhanced Advanced LIGO Detectors in the Era of Gravitational-Wave Astronomy, 
Phys. Rev. Lett. {\bf 123}, 231107 (2019)

\bibitem{ArcenesePRL2019}
F. Acernese {\it et al.} (Virgo Collaboration),
Increasing the Astrophysical Reach of the Advanced Virgo Detector via the Application of Squeezed Vacuum States of Light, 
Phys. Rev. Lett. {\bf 123}, 231108 (2019)

\bibitem{LudlowRMP2015}
A.D. Ludlow, M.M. Boyd, J. Ye, E. Peik and P.O. Schmidt,
Optical atomic clocks,
Rev. Mod. Phys. {\bf 87}, 637 (2015)

\bibitem{RondinRPP2014}
L. Rondin, J.P. Tetienne, T. Hingant, J.F. Roch, P. Maletinsky and V. Jacques,
Magnetometry with nitrogen-vacancy defects in diamond,
Rep. Prog. Phys. {\bf 77}, 056503 (2014)

\bibitem{CroninRMP2009}
A.D. Cronin, J. Schmiedmayer, and D.E. Pritchard, 
Optics and interferometry with atoms and molecules,
Rev. Mod. Phys. {\bf 81}, 1051 (2009)

\bibitem{BarrettPS2016}
B. Barrett, A. Bertoldi and P. Bouyer,
Inertial quantum sensors using light and matter, 
Phys. Scr. {\bf 91} 053006 (2016).

\bibitem{TaylorPR2016}
M. Taylor and W. Bowen,
Quantum metrology and its application in biology,
Physics Reports {\bf 615}, 1-59 (2016)

\bibitem{LanePRA1993}
A.S. Lane, S.L. Braunstein and C.M. Caves, 
Maximum-likelihood statistics of multiple quantum phase measurements, 
Phys. Rev. A {\bf 47}, 1667 (1993).

\bibitem{PezzePRL2007}
L. Pezz\`e, A. Smerzi, G. Khoury, J.F. Hodelin and D. Bouwmeester, 
Phase Detection at the Quantum Limit with Multiphoton Mach-Zehnder Interferometry,
Phys. Rev. Lett. {\bf 99}, 223602 (2007)

\bibitem{OlivaresJPB2009}
S. Olivares and M.G. Paris, 
Bayesian estimation in homodyne interferometry, 
J. Phys. B: At. Mol. Opt. Phys. {\bf 42}, 055506 (2009).

\bibitem{KrischekPRL2011}
R. Krischek, C. Schwemmer, W. Wieczorek, H. Weinfurter, P. Hyllus, L. Pezz\`e and A. Smerzi, 
Useful Multiparticle Entanglement and Sub-Shot-Noise Sensitivity in Experimental Phase Estimation,
Phys. Rev. Lett. {\bf 107}, 080504 (2011)

\bibitem{XiangNATPHOT2011}
G.Y. Xiang, B.L. Higgins, D.W. Berry, H.M. Wiseman and G.J. Pryde,
Entanglement-enhanced measurement of a
completely unknown optical phase
Nat. Photonics {\bf 5}, 43 (2011).

\bibitem{PezzeBOOK2014}
L. Pezz\`e and A. Smerzi, 
Quantum Theory of Phase Estimation, in Atom Interferometry, Proceedings of the International School of Physics "Enrico Fermi", Course 188, Varenna, edited by G. M. Tino and M. A. Kasevich
(IOS Press, Amsterdam, 2014), p. 691;
arXiv:1411.5164.

\bibitem{LiENTROPY2018}
Y. Li, L. Pezz\`e, M. Gessner, Z. Ren, W. Li and A. Smerzi,  
Frequentist and Bayesian Quantum Phase Estimation,
Entropy {\bf 20}, 628 (2018)

\bibitem{RubioJPC2018}
J. Rubio, P. Knott, and J. Dunningham, 
Non-asymptotic analysis of quantum metrology protocols beyond the Cram\'er–Rao bound,
J. Phys. Commun. 2 015027 (2018). 

\bibitem{CiminiPRA2020}
V Cimini, MG Genoni, I Gianani, N Spagnolo, F Sciarrino, M Barbieri,
Diagnosing imperfections in quantum sensors via generalized Cramér-Rao bounds,
Phys. Rev. Appl. {\bf 13}, 024048 (2020).

\bibitem{KayBOOK}
S.M. Kay,
{\it Fundamentals of Statistical Signal Processing: Estimation Theory, Volume I} (Prentice Hall: Upper Saddle River, NJ, USA, 1993).

\bibitem{LehmannBOOK}
E.L. Lehmann, and G. Casella,
{\it Theory of Point Estimation},
Springer Texts in Statistics (Springer, New York, 1998).

\bibitem{VanTreesBOOK}
H.L. Van Trees, and K.L. Bell (eds.), {\it Bayesian Bounds for Parameter Estimation and Nonlinear Filtering/Tracking}
(Wiley: New York, NY, USA, 2007).


\bibitem{WiebePRL2016}
N. Wiebe and C. Granade,
Efficient Bayesian Phase Estimation, 
Phys. Rev. Lett. {\bf 117}, 010503 (2016).

\bibitem{PaesaniPRL2017}
S. Paesani, A.A. Gentile, R. Santagati, J. Wang, N. Wiebe, D.P. Tew, J.L. O'Brien and M.G. Thompson, 
Experimental Bayesian Quantum Phase Estimation on a Silicon Photonic Chip, 
Phys. Rev. Lett. {\bf 118}, 100503 (2017).

\bibitem{SantagatiPRX2019}
%R. Santagati, A.A. Gentile, S. Knauer, S. Schmitt, S. Paesani, C. Granade, N. Wiebe, C. Osterkamp, L.P. McGuinness, J. Wang, M.G. Thompson, J.G. Rarity, F. Jelezko, and A. Laing,
R. Santagati {\it et al.}
Magnetic-Field Learning Using a Single Electronic Spin in Diamond with One-Photon Readout at Room Temperature,
Phys. Rev. X {\bf 9}, 021019 (2019)

\bibitem{BerryPRL2000}
D.W. Berry and H.M. Wiseman,
Optimal states and almost optimal adaptive measurements for quantum interferometry,
Phys. Rev. Lett. {\bf 85}, 5098 (2000).

\bibitem{HiggingNATURE2007}
B.L. Higgins, D.W. Berry, S.D. Bartlett, H.M. Wiseman and G.J. Pryde, 
Entanglement-free Heisenberg-limited phase estimation, 
Nature {\bf 450}, 393-396 (2007).

\bibitem{BerniNATPHOT2015}
A.A. Berni, T. Gehring, B.M. Nielsen, V. Händchen, M.G. Paris and U.L. Andersen, 
%A.A. Berni {\it et al.}
Ab initio quantum-enhanced optical phase estimation using real-time feedback control,
Nat. Photonics {\bf 9}, 577-581 (2015).

\bibitem{BonatoNATNANO2015}
C. Bonato, M.S. Blok, H.T. Dinani, D.W. Berry, M.L. Markham, D.J. Twitchen and R. Hanson,
%C. Bonato {\it et al.}
Optimized quantum sensing with a single electron spin using real-time adaptive measurements, 
Nat. Nanotechnology {\bf 11}, 247-252 (2015).

\bibitem{MartinezNJP2016}
F. Mart\'{\i}nez-Garc\'{\i}a, D. Vodola and M. M\"uller,
Adaptive Bayesian phase estimation for quantum error correcting codes,
New J. Phys. {\bf 21}, 123027 (2019).

\bibitem{HincksNJP2012}
I. Hincks, C. Granade and D.G. Cory,
Statistical inference with quantum measurements:methodologies for nitrogen vacancy centers in diamond,
New J. Phys. {\bf 20}, 013022 (2012).

\bibitem{AhronSR2019}
N. Aharon, A. Rotem, L.P. McGuinness, F. Jelezko, A. Retzker and Z. Ringel,
%N. Aharon {\it et al.}
NV center based nano-NMR enhanced by deep learning,
Scientific Reports {\bf 9}, 17802 (2019) 

\bibitem{SchwartzSR2019}
L. Schwartz, J. Rosskopf, S. Schmitt, B. Tratzmiller, Q. Chen, L.P. McGuinness, F. Jelezko and M.B. Plenio,
%L. Schwartz {\it et al.}
Blueprint for nanoscale NMR,
Scientific Reports {\bf 9}, 6938 (2019) 

\bibitem{NielsenBOOK}
M.A. Nielsen, {\it Neural Networks and Deep Learning}
(Determination Press, 2015), available at 
http://neuralnetworksanddeeplearning.com

\bibitem{MurphyBOOK}
K.P. Murphy, {\it Machine Learning: A Probabilistic Perspective}
(MIT Press, Cambridge, MA, 2012)

\bibitem{MethaPR2019}
P. Metha, M. Bukov, C.H. Wang, A.G.R. Day, C. Richardson, C.K. Fisher and D.J. Schwab, 
%P. Metha {\it et al.}
High-bias, low-variance introduction to Machine Learning for physicists,
Physics Reports {\bf 810} 1-124 (2019)

\bibitem{lecun2010mnist}
Y. LeCun, C. Cortes and C. Burges,
(ATT Labs [Online], 2010), available at 
http://yann.lecun.com/exdb/mnist

\bibitem{DunjkoRPP2018}
V. Dunjko and H.J. Briegel,
Machine learning \& artificial intelligence in the quantum domain: 
a review of recent progress, 
Rep. Prog. Phys. {\bf 81}, 074001 (2018)

\bibitem{CarleoRMP2019}
G. Carleo, I. Cirac, K. Cranmer, L. Daudet, M. Schuld, N. Tishby, L. Vogt-Maranto and L. Zdeborov\'{a},
%G. Carleo {\it et al.}
Machine learning and the physical sciences,
Rev. Mod. Phys. {\bf 91}, 045002 (2019)

\bibitem{PolinoARXIV}
E. Polino, M. Valeri, N. Spagnolo and F. Sciarrino,
Photonic Quantum Metrology, 
AVS Quantum Sci. {\bf 2}, 024703 (2020). 

\bibitem{HentschelPRL2010}
A. Hentschel and B.C. Sanders, 
Machine Learning for Precise Quantum Measurements, 
Phys. Rev. Lett. {\bf 104}, 063603 (2010)

\bibitem{HentschelPRL2011}
A. Hentschel and B.C. Sanders, 
Efficient Algorithm for Optimizing Adaptive Quantum Metrology Process, 
Phys. Rev. Lett. {\bf 107}, 233601 (2011)

\bibitem{LovettPRL2013}
N.B. Lovett, C. Crosnier, M. Perarnau-Llobet and B.C. Sanders,
Differential Evolution for Many-Particle Adaptive Quantum Metrology,
Phys. Rev. Lett. {\bf 110}, 220501 (2013)

\bibitem{LuminoPRAPP2018}
A. Lumino, E. Polino, A.S. Rab, G. Milani, N. Spagnolo, N. Wiebe and F. Sciarrino,
%A. Lumino {\it et al.}
Experimental Phase Estimation Enhanced by Machine Learning,
Phys. Rev. Appl. {\bf 10}, 044033 (2018)

\bibitem{XiaoSR2019}
T. Xiao, J. Huang, J. Fan and G. Zeng, 
Continuous-variable Quantum Phase Estimation based on Machine Learning,
Scientific Reports {\bf 9}, 12410 (2019) 

\bibitem{XuNPJ2019}
H. Xu, J. Li, L. Liu, Y. Wang, H. Yuan and X. Wang,
%H. Xu {\it et al.}
Generalizable control for quantum parameter estimation through reinforcement learning,
npj Quantum Information {\bf 9}, 82 (2019)

\bibitem{PalittapongarnpimPRA2019}
P. Palittapongarnpim and B. Sanders,
Robustness of quantum-enhanced adaptive phase estimation, 
Phys. Rev. A {\bf 100}, 012106 (2019).

\bibitem{PengPRA2020}
Y. Peng and H. Fan, 
Feedback ansatz for adaptive-feedback quantum metrology training with machine learning, 
Phys. Rev. A {\bf 101}, 022107 (2020).

\bibitem{SchuffNJP2020}
J. Schuff, L.J. Fiderer and D. Braun, 
Improving the dynamics of quantum sensors with reinforcement learning, 
New J. Phys. {\bf 22}, 035001 (2020).

\bibitem{FidererARXIV2020}
L.J. Fiderer, J. Schuff and D. Braun,
Neural-Network Heuristics for Adaptive Bayesian Quantum Estimation,
PRX Quantum {\bf 2}, 020303 (2021).

\bibitem{QianAppPhysLett2021}
P. Qian, X. Lin, F. Zhou, R. Ye, Y. Ji, B. Chena, G. Xie, and Nanyang Xua,
%P. Qian {\it et al.}
Machine-learning-assisted electron-spin readout of nitrogen-vacancy center in diamond,
Appl. Phys. Lett.  {\bf 118}, 084001 (2021)

\bibitem{HainePRL2012}
S. Haine and J. Hope,
A Machine-Designed Sensor to Make Optimal Use of Entanglement-Generating Dynamics for Quantum Sensing,
Phys. Rev. Lett. {\bf 124}, 060402 (2020)

\bibitem{GrossPRL2010}
D. Gross, Y.K. Liu, S.T. Flammia, S. Becker and J. Eisert,
Quantum State Tomography via Compressed Sensing
Phys. Rev. Lett. {\bf 105}, 150401 (2010)

\bibitem{XuARXIV2018}
Q. Xu and S. Xu,
Neural network state estimation for full quantum state tomography
{arXiv:1811.06654} (2018)

\bibitem{TorlaiNATPHYS2018}
G. Torlai, G. Mazzola, J. Carrasquilla, M. Troyer, R. Melko and G. Carleo,
%G. Torlai {\it et al.}
Neural-network quantum state tomography,
Nature Physics {\bf 14}, 447-450 (2018)

\bibitem{QuekARXIV}
Y. Quek, S. Fort, H.K. Ng
Adaptive Quantum State Tomography with Neural Networks, 
arXiv:1812.06693. 

\bibitem{XinNPJ2019}
T. Xin, S. Lu, N. Cao, G. Anikeeva, D. Lu, J. Li, G. Long and B. Zeng,
%T. Xin {\it et al.}
Local-measurement-based quantum state tomography via neural networks,
npj Quantum Information {\bf 14}, 109 (2019)

\bibitem{CarrasquillaNATMI2019}
J. Carrasquilla, G. Torlai, R.G. Melko and L. Aolita,
Reconstructing quantum states with generative models,
Nature Machine Intelligence {\bf 1}, 155 (2019)

\bibitem{MacaronePalmieriNPJ2020}
A. Macarone Palmieri, E. Kovlakov, F. Bianchi, D. Yudin, S. Straupe, J.D. Biamonte and S. Kulik,
%A. Macarone Palmieri {\it et al.}
Experimental neural network enhanced quantum tomography,
npj Quantum Inf.ormation {\bf 6}, 20 (2020) 

\bibitem{SpagnoloSCIREP2017}
N. Spagnolo, E. Maiorino, C. Vitelli, M. Bentivegna, A. Crespi, R. Ramponi, P. Mataloni, R. Osellame and F. Sciarrino,
%N. Spagnolo {\it et al.}
Learning an unknown transformation via a genetic approach,
Scientific Reports {\bf 7}, 14316 (2017)

\bibitem{RocchettoSCIADV2019}
A. Rocchetto, S. Aaronson, S. Severini, G. Carvacho, D. Poderini, I. Agresti, M. Bentivegna and F. Sciarrino,
%A. Rocchetto {\it et al.}
Experimental learning of quantum states,
Science Advances {\bf 5}, 1946 (2019)

\bibitem{Yu2019}
%S. Yu, F. Albarrán-Arriagada, J.C. Retamal, Y.T. Wang, W. Liu, Z.J. Ke, Y. Meng, Z.P. Li, J.S. Tang, E. Solano, L. Lamata, C.F. Li and G.C. Guo,
S. Yu {\it et al.}
Reconstruction of a Photonic Qubit State with Reinforcement Learning,
Adv. Q. Tech. {\bf 2}, 1800074 (2019)

\bibitem{Aaronson2019}
S. Aaronson,
The learnability of quantum states,
Proc. R. Soc. A. {\bf 463}, 3089 (2007)

\bibitem{TorlaiARXIV2019}
G. Torlai, G. Mazzola, G. Carleo and A. Mezzacapo,
Precise measurement of quantum observables with neural-network estimators,
{arXiv:1910.07596} (2019)

\bibitem{FlurinPRX2020}
E. Flurin, L.S. Martin, S. Hacohen-Gourgy and I. Siddiqi, 
Using a Recurrent Neural Network to Reconstruct Quantum Dynamics of a Superconducting Qubit from Physical Observations,
Phys. Rev. X {\bf 10}, 011006 (2020)

\bibitem{GranadeNJP2012}
C. E. Granade, C. Ferrie, N. Wiebe and D.G. Cory,
Robust online Hamiltonian learning,
New J. Phys. {\bf 14}, 103013 (2012).

\bibitem{WangNATPHYS2017}
J. Wang, S. Paesani, R. Santagati, S. Knauer, A.A. Gentile, N. Wiebe, M. Petruzzella, J.L. O'Brien, J.G. Rarity, A.Laing and M.G. Thompson,
%J. Wang {\it et al.}
Experimental quantum Hamiltonian learning,
Nature Physics {\bf 13}, 551 (2017)
    
\bibitem{WangARXIV2019}
D. Wang, S. Wei, A. Yuan, F. Tian, K. Cao, Q. Zhao, D. Xue and S. Yang,
%D. Wang {\it et al.}
Machine learning magnetic parameters from spin configurations,
{arXiv:1910.05829} (2019)   

\bibitem{WozniakowskiARXIV2020}
A. Wozniakowski, J. Thompson, M. Gu and F. Binder,
Boosting on the shoulders of giants in quantum device calibration,
{arXiv:2005.06194} (2020)   

\bibitem{YouARXIV2019}
%C. You, N. Bhusal, A. Lambert, C. Dong, A. Perez-Leija, R.J. Leon-Montiel, A. Javaid and O.S. Magana-Loaiza, 
C. You {\it et al.}
Identification of Light Sources using Artificial Neural Networks,
{arXiv:1909.08060} (2019)

\bibitem{GebhartPRR2020}
V. Gebhart and M. Bohmann, 
Neural-network approach for identifying nonclassicality from click-counting data,
Phys. Rev. Research {\bf 2}, 023150 (2020) 

\bibitem{GreplovaARXIV2017}
E. Greplova, C.K. Andersen and K. M\o{}lmer, 
Quantum parameter estimation with a neural network,
{arXiv:1711.05238} (2017)

\bibitem{LiuJPHYSB2019}
W. Liu, J. Huang, Y. Li, H. Li, C. Fang, Y. Yu and G. Zeng,
%W. Liu {\it et al.}
Parameter estimation via weak measurement with machine learning,
J. Phys. B: Atomic, Molecular and Optical Physics {\bf 52}, 045504 (2019).

\bibitem{KhanahmadiARXIV}
M. Khanahmadi and K. M\o{}lmer, 
Time-dependent atomic magnetometry with a recurrent neural network,
Phys. Rev. A {\bf 103}, 032406 (2021) 

\bibitem{CiminiPRL2019}
V. Cimini, I. Gianani, N. Spangolo, F. Leccese, F. Sciarrino and M. Barbieri,
%V. Cimini {\it et al.}
Calibration of Quantum Sensors by Neural Networks, 
Phys. Rev. Lett. {\bf 123}, 230502 (2019)

\bibitem{BraunsteinPRL1994}
S.L. Braunstein and C.M. Caves, 
Statistical Distance and the Geometry of Quantum States,
Phys. Rev. Lett. {\bf 72}, 3439 (1994).

\bibitem{KitagawaPRA1993}
M. Kitagawa and M. Ueda, 
Squeezed spin states,
Phys. Rev. A {\bf 47}, 5138 (1993)

\bibitem{PezzePRL2009}
L. Pezz\`e and A. Smerzi, 
Entanglement, Nonlinear Dynamics, and the Heisenberg Limit, 
Phys. Rev. Lett. {\bf 102}, 100401 (2009). 

\bibitem{chollet2015keras}
F. Chollet, et al., {\it Keras}
(2015), available at 
http://keras.io

\bibitem{KingmaARXIV2014}
D.P. Kingma and J. Ba, 
Adam: A Method for Stochastic Optimization,
{arXiv:1412.6980} (2014)



%%%%%%%%%%%%%%%%%%%%%%%%%%%%%%%%%%%%%%%%%%%%%%%%%%%%%%%%%%%%%
\end{thebibliography}
\end{document}